\documentclass[12pt]{iopart}
\usepackage{graphicx}
\usepackage{epstopdf}
\usepackage{color}
\usepackage{amsfonts}
\usepackage{subfigure}
\usepackage{bm}

\begin{document}

\title[Seeded Rydberg excitation]{Seeded excitation avalanches in off-resonantly driven Rydberg gases}

\author{C. Simonelli, M.M. Valado, G. Masella, L. Asteria, E. Arimondo, D. Ciampini, O. Morsch}
\address{Dipartimento di Fisica ``E. Fermi'', Universit\`a di Pisa, Largo Bruno Pontecorvo 3, 56127 Pisa, Italy}
\address{INO-CNR, Via G. Moruzzi 1, 56124 Pisa, Italy}


\ead{morsch@df.unipi.it}

\begin{abstract}
We report an experimental investigation of the facilitated excitation dynamics in off-resonantly driven Rydberg gases by separating the initial off-resonant excitation phase from the facilitation phase, in which successive facilitation events lead to excitation avalanches. We achieve this by creating a controlled number of initial seed excitations. Greater insight into the avalanche mechanism is obtained from an analysis of the full counting distributions. We also present simple mathematical models and numerical simulations of the excitation avalanches that agree well with our experimental results.

\end{abstract}

\pacs{34.20.Cf, 32.80.Ee}
\vspace{2pc}
\noindent{\it Keywords}: Rydberg excitation, quantum control\\
\vspace{2pc}
\submitto{\JPB}
%
%
%

\section{Introduction}
Rydberg excitations in ultracold atoms have become a rich and versatile model system for strongly correlated phenomena in recent years~\cite{Bloch:08}. The strong van der Waals and dipole-dipole interactions between high-lying Rydberg states give rise to a host of phenomena such as the dipole blockade, in which the creation of more than one Rydberg excitation within a certain radius is strongly suppressed ~\cite{ Urban:09, Gaetan:09, Beguin:13,Petrosyan:2013}, and facilitated excitation~\cite{Lesanovsky:14, Ates:07b}. The latter arises for off-resonant excitation of the atomic sample and relies on the Rydberg-Rydberg interaction that shifts the energy levels of the two atoms and thus exactly compensates the laser detuning. The laser detuning, therefore, represents an experimental handle for the selective excitation of pairs of atoms at a well-defined interatomic distance. In recent experiments signs of this facilitation have been observed in the form of lineshape asymmetries  ~\cite{Carr:13, Malossi:14}, the formation of excitation clusters  ~\cite{Garttner:13} deduced from an analysis of the full counting statistics  ~\cite{Schempp:14, Malossi:14} and the transmission properties of a room-temperature gas  ~\cite{Loew:12,Urvoy:15}.\\
\indent In typical experiments involving facilitated excitation~\cite{Lesanovsky:14, Ates:07b,Schempp:14, Malossi:14,Loew:12, Urvoy:15}, the off-resonant excitation dynamics is triggered by the presence within the atomic cloud of one (or more) atoms already excited to the target Rydberg level.
In order to explore and exploit the facilitation process we have developed a 'seed' technique in which a resonant laser pulse creates the required initial excitation(s), and the successive facilitation process is mediated by a second, off-resonant laser pulse. The temporal and spatial control over the laser pulses and the atomic cloud provides a promising scheme for the controlled preparation of strongly correlated atomic systems.
Here we report a detailed investigation of the facilitation mechanism based on the seed technique, which allows us to separate the initial off-resonant excitation phase from the facilitation phase in which successive facilitated excitation events lead to avalanche-like chain reactions. Using the full counting distributions of the number of excitations we study the avalanche phase and compare our results to simple mathematical models and numerical simulations.\\
    \indent In this work first briefly recap the basic idea of the facilitation mechanism and extend it to non-uniform atomic density distributions. We then process to describe our experimental apparatus and, in particular, the different laser geometries determining the spatial distribution of the facilitated excitations, as well as the laser configuration used for creating the seed excitations. Experimental results on the dynamics of the seeded avalanche excitations and the full-counting distributions in different excitation regimes are presented in Section 4. Section 5 describes two different models predicting the time dependence and the probability distributions of the excitation numbers. The numerical simulations presented predict full counting distributions that are in good agreement with the experimental results in different parameter regimes.

\section{The facilitation mechanism}
In the dipole blockade picture the strong interaction between Rydberg atoms means that an already excited Rydberg atom can suppress the
excitation of a second one at a distance smaller than the so-called blockade radius. In the case of coherent excitation all the atoms inside a blockade volume share one collective excitation.

\begin{figure}[htbp]
\begin{center}
\includegraphics[width=9 cm]{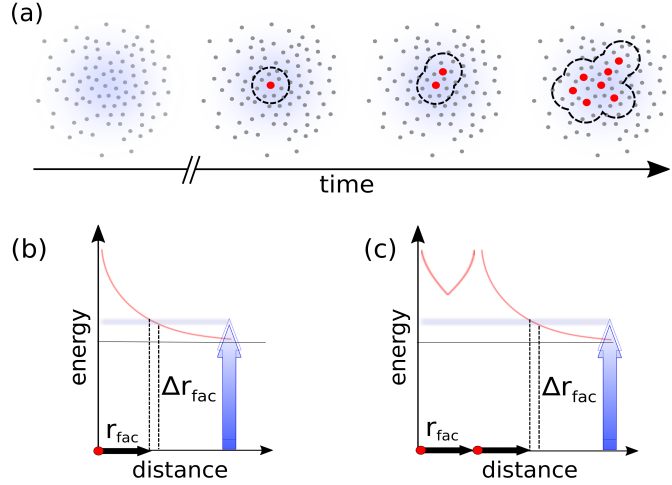}
\caption{(a) Schematic representation of the facilitation process and the seeded avalanche dynamics. Initially, the off-resonant creation of single Rydberg excitations is unlikely and the number of excitations in the system is close to zero. After creating a seed excitation in the atomic cloud, facilitation leads to a chain reaction of excitations that eventually fill the entire cloud. (b) Energy scheme of the first facilitated excitation and (c) of the successive facilitated excitations.}
\label{Seed}
\end{center}
\end{figure}

If, on the other hand, we consider off-resonant Rydberg excitation, the van der Waals interaction can
compensate the energy mismatch, thus facilitating the excitation of
pairs of Rydberg atoms at an appropriate distance. Two atoms at a distance $r$ from each other can be
simultaneously coherently excited if $r$ and the detuning $\Delta$ of the excitation laser fulfil the following
condition ~\cite{Weimer:2015}:
\begin{equation}
r = \left(\frac{C_6}{2 \hbar \Delta}\right)^{\frac{1}{6}}
\end{equation}
A different process occurs if a Rydberg excitation is already present in the system: the van der Waals interaction can then facilitate successive Rydberg excitations of ground state atoms. In Fig.~\ref{Seed} a scheme of this
off-resonant excitation is presented. The Rydberg excitation already present in the system shifts the energy levels of atoms located inside  a facilitation shell of radius $r_{fac }$ and thickness $\delta r_{fac }$~\cite{Lesanovsky:14} into resonance with the Rydberg transition, where
\begin{eqnarray}
r_{fac}&= \left(\frac{C_6}{\hbar \Delta}\right)^{\frac{1}{6}}, \nonumber \\
\delta r_{fac }&= r_{fac} \frac{\gamma}{6 \Delta},
\label{facilradii}
\end{eqnarray}
 with $C_6$  the van der Waals coefficient  $ (h\times 869.7\,\mathrm{GHz}\times \mu \mathrm{m^6}$ for the 70S state \cite{WalkerSaffman:08} used in our experiments), and $\gamma$ the combined laser and residual Dopper linewidth (around $2\pi \times 0.7\,\mathrm{MHz}$ in our case). Once an excitation is created in the facilitation shell, it creates a second facilitation shell around it in which further atoms can be excited. In this way, a chain reaction or 'avalanche' ensues that stops when there are no more atoms inside the next facilitation shell.\\
\indent  If $N_g$ denotes the mean number of atoms within the facilitation shell, the probability $P$ of there being at least one atom in the shell can be calculated (assuming a Poissonian process) as
\begin{equation}
P( N_g)=1-e^{-\langle N_g\rangle}.
\end{equation}
In a 3D geometry, for a  mean atomic density $\overline{\rho}$  at the shell location, the mean number of atoms in the shell is given by
\begin{equation}
\langle N_g( r_{fac})\rangle = 4 \pi r_{fac}^2 \delta r_{fac }\overline{\rho}.
\end{equation}
In a 1D geometry along the $x$ axis, with the atoms confined transversely within an area $\sigma$, and with a spatially dependent density  $\rho(x)$, the mean atom number in the shell is given by
\begin{equation}
N_g( x_{fac}) = 2 \sigma \delta x_{fac } \rho(x).
\label{1Dfacilitation}
\end{equation}
In the typical case of a Gaussian atomic density distribution, which is a good approximation for the density distribution inside a magneto-optical trap, as the facilitation process moves outwards from the centre (where the creation of the initial seed excitation is most likely), the density in the next facilitation shell decreases, and eventually the facilitation process reaches a stage where the probability of there being at least one atom in the next shell is close to zero. At that stage, denoted as  the 'saturation stage', the number of Rydberg excitations will reach its saturation limit.\\

\section{Experimental apparatus}

\begin{figure}[htbp]
\begin{center}
\includegraphics[width=9 cm]{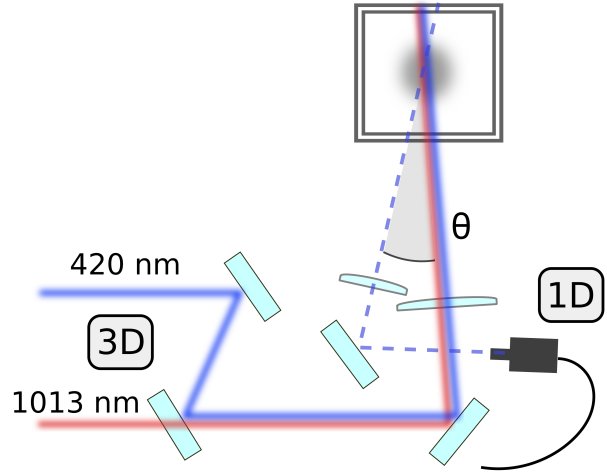}
\caption{
Sketch of the optical alignment for the quasi 1D and 3D geometries.  The 3D geometry is realized with co-propagating 1013 nm and 420 nm beams, while for the quasi 1D geometry a tightly focused 420 nm beam intersects the 1013 nm beam at an angle $\theta$.}
\label{SeedCreation}
\end{center}
\end{figure}

Our experiments are performed in magneto-optical traps (MOTs) of $^{87}$Rb atoms with typical sizes around $150\,\mathrm{\mu m}$  and peak densities $\rho_0$ of up to $10^{11}$ cm$^{-3}$ (for details see ~\cite{Viteau:13,Malossi:14}). With the MOT beams switched off,  laser beams at 420 nm and 1013 nm excite atoms to the $70\mathrm{S}$ Rydberg state through a two-photon process via the $6\,\mathrm{P}_{3/2}$ intermediate state. The 420 nm laser is detuned by $\approx 1\,\mathrm {GHz} $ from that level. We usse laser configurations realizing different excitation volumes, using the two lasers in copropagating or angle configuration. A quasi 1D geometry is realized with the 420 nm beam focused to around $6 \,\mathrm{\mu m}$, comparable to the facilitation radius, while the laser beam at 1013 nm has a waist of $110\,\mathrm{\mu m}$. A 3D geometry is realized when the 420 nm laser has a waist of 40 $\mu$m and co-propagates with the 1013 nm laser. In the following, the effective excitation volumes and the number of atoms will be reported in the captions of the experimental figures.\\
\indent At the end of an excitation sequence the Rydberg atoms are field ionized and the ions thus produced are accelerated to a channeltron. The overall $\eta$ detection efficiency of our system is around 40$\%$. In the following we report the measured number $N_{obs}$ of ions, from which the actual number of Rydberg excitations $N$ is calculated as $N=N_{obs}/\eta$.\\
\indent The resonant two-photon Rabi frequency for the Rydberg excitation is $\Omega/(2\pi)=35 \pm 7 \,\mathrm{kHz}$. Since this Rabi frequency is smaller than the laser linewidth $\gamma$, the (resonant) dynamics of our system takes place in the incoherent regime characterized by a transition rate $\Gamma = \Omega^2/{2\gamma}$ \cite{Lesanovsky:13}. \\
\indent In order to create initial seed excitations we use either the 1013 nm beam (derived from the plus first diffraction order of the acousto-optic modulator used to control the pulses) with a shifted frequency or a separate resonant beam at 1013 nm. In the latter case, as in Fig.~\ref{SeedCreation}, the resonant pulse is created with the -1 diffraction order of the acousto-optic modulator, while the zeroth order off-resonant excitation beam is constantly present. The pulse timing in this case is achieved solely with the 420 nm beam. We verified that the constant presence of the off-resonant 1013 nm laser radiation alone has no effect on the atoms. The frequency-switched single beam technique works for two-photon detuning $\Delta/2\pi$ up to $30 \,\mathrm{MHz}$, whereas the two-beam technique is used for $\Delta/2\pi$ between 55 and 80 MHz. These limits stem from the diffraction efficiency of the acousto-optic modulator at different frequencies and from the variation of the diffraction angle with frequency that leads to a shift in the beam position with respect to the atomic cloud.\\
\indent  The (effective) number of ground-state atoms in our experiment, and hence the density of the clouds (keeping their size and shape constant), is controlled by depumping a fraction of the atoms from the F=2 hyperfine ground state (which is coupled to the Rydberg state by the excitation lasers) into the F=1 state (which lies $6.8\,\mathrm{GHz}$ below the F=2 state is is, therefore, effectively decoupled from the Rydberg transition) using a resonant laser pulse of duration up to 2 $\mu$s, as presented in~\cite{Valado:15}.

\section{Experimental results}
\begin{figure}[htbp]
\begin{center}
\includegraphics[width=9 cm]{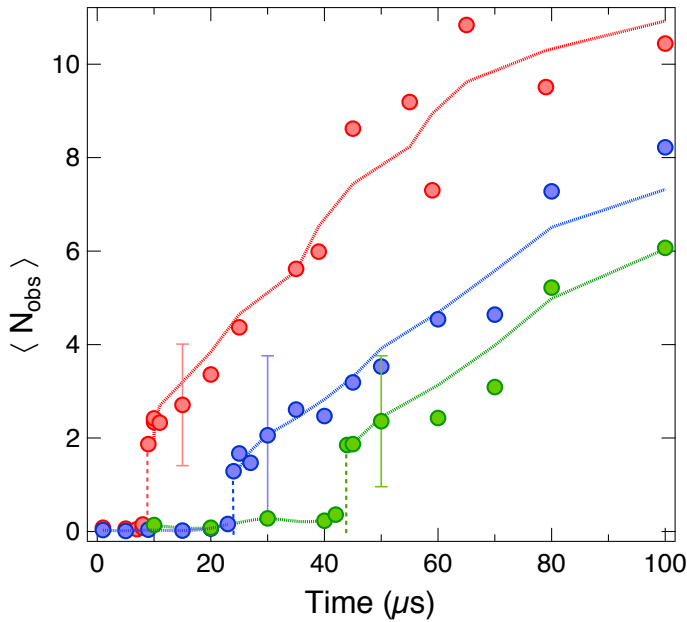}
\caption{Mean number $\langle N_{obs}\rangle$ of Rydberg excitations as a function of time for off-resonant excitation at $\Delta/2\pi =+75\,\mathrm{MHz}$ and different values of the time $t_ {seed}$ at which $\approx 2$ seeds are created by a short resonant pulse. The red data points correspond to $t_ {seed}=10\,\mathrm{\mu s}$, blue data to $t_ {seed}=25\,\mathrm{\mu s}$ and green data to $45\,\mathrm{\mu s}$. The experiments were carried out in a 3D geometry with  $N=1,700,000$ atoms (of which $210,000$ in the interaction volume) in a MOT with dimensions $160\,\mathrm{\mu m} \times 130\,\mathrm{\mu m}\times 100\,\mathrm{\mu m}$.}
\label{Seedresults}
\end{center}
\end{figure}

The seeded Rydberg excitation is based on the application at time $t_ {seed}$ of a short (around $0.5\,\mathrm{\mu s}$) resonant pulse ($\Delta=0$). That pulse creates a controllable number of Rydberg excitations (seeds) limited by the interaction time and, in the regime of large numbers of seeds, by the dipole blockade.  Subsequently, an off-resonant laser is applied which induces the facilitated excitation process described above. Figure \ref{Seedresults} reports experimental results for this in a 3D geometry: at time $t_ {seed}$ the resonant pulse creates around $2$ seed excitations, after which the excitation laser is switched to $\Delta/2\pi=+75\,\mathrm{MHz}$. Typical results  for $t_ {seed}=10\,\mathrm{\mu s}$, $25\,\mathrm{\mu s}$ and $45\,\mathrm{\mu s}$ are shown. Between $t=0$ and $t=t_ {seed}$ the mean number of excitations remains close to zero and grows very slowly due to the large detuning of the excitation laser. After the injection of seeds at $t_ {seed}$, the rate of excitation rises sharply and the mean number continues to increase up to $\langle N_{obs}\rangle\approx 10$, at which point saturation sets in.  As can be clearly seen in Fig.  \ref{Seedresults}, the atomic dynamics after the injection of the seed is largely independent of $t_ {seed}$, as one might expect.

\begin{figure}[htbp]
\begin{center}
\includegraphics[width=8.5cm]{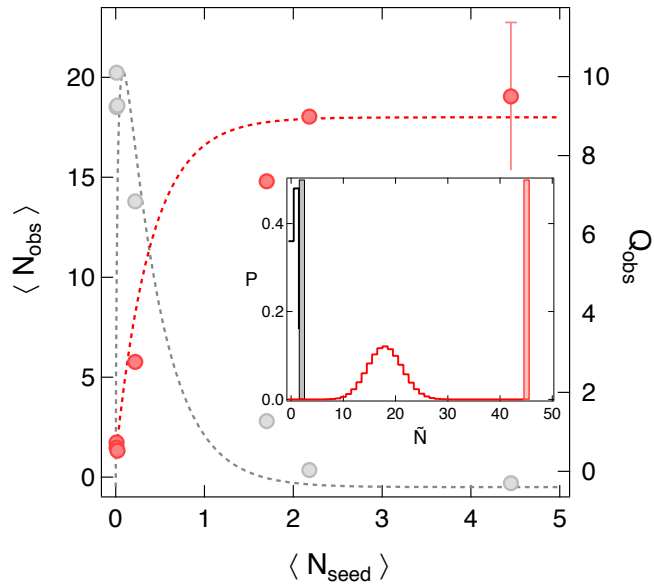}
\caption{Off-resonant excitation in a 3D geometry at $\Delta/2\pi =+27\,\mathrm{MHz}$. The mean number $\langle N_{obs}\rangle$ of Rydberg excitations (red circles and left axis) and the Mandel Q-parameter  $Q_{obs}$ (grey circles and right axis) after $100\,\mathrm{\mu s}$ excitation are plotted vs the measured number of seed excitations $\langle N_{seed}\rangle$. As $\langle N_{seed}\rangle$ increases,  the system reaches saturation. The dashed lines are the predictions of the bimodal model with $N_1=2$ and $N_2=45$.  The inset reports the corresponding probability distribution $P(\tilde{N})$ of the bimodal model excluding (vertical bars) and including (stepwise line) the binomial convolution of ref.~\cite{Malossi:14} taking into account the  $\eta$ detection efficiency. The values $\tilde{N}$ correspond to $N_{obs}$ for the line and $N$ for the bars. The experiments were performed in a 3D geometry with  $N=90,000$ atoms (of which $50,000$ in the interaction volume) in a MOT with dimensions $80\,\mathrm{\mu m} \times 40\,\mathrm{\mu m}\times 60\,\mathrm{\mu m}$.}
\label{Bimodal}
\end{center}
\end{figure}

In order to confirm our interpretation of the role of the seed excitation, we perform an experiment in the 3D geometry in which a variable number $\langle N_{seed}\rangle$ of seed excitations is created at $t=0$ and the system is thereafter off-resonantly excited for $100\,\mathrm{\mu s}$. The results of this experiment are reported in Fig. \ref{Bimodal}. The mean number of seeds $\langle N_{seed}\rangle$ created at the beginning of the dynamics is varied between 0 and around $5$ by changing the resonant pulse duration. From the results shown in Fig. \ref{Seedresults} one might expect that each time at least one seed is created at $t=0$, a facilitation avalanche is triggered and at $100\,\mathrm{\mu s}$ the system has reached the saturation regime. This behaviour is confirmed by the experimental data of Fig. \ref{Bimodal}. From that data we also derive the (observed) Mandel Q-parameter  $Q_{obs}$ defined as~\cite{Mandel:1982, Viteau:2012}
\begin{equation}
Q_{obs}=\frac{\langle (N_{obs}-\langle N_{obs}\rangle)^2\rangle}{\langle N_{obs}\rangle}-1.
\end{equation}
The dependence of $Q_{obs}$ on the mean number of seeds is reported in Fig.~\ref{Bimodal}.  For small values of $\langle N_{seed}\rangle$ the fluctuations in the mean number of excitations at $100\,\mathrm{\mu s}$ are large as the system will sometimes (when a seed is created at $t=0$) end up with a large number of excitations, and sometimes with very few. As the mean number of seeds (and hence the probability of creating at least one seed) grows, the $Q$-parameter decreases and becomes slightly negative, indicating a sub-Poissonian distribution that is compatible with the interpretation of almost deterministically triggering an avalanche that always results in the same final number of excitations. \\
\indent While in Fig. 4 the (small) mean number of seeds essentially determined only the probability of starting the facilitation avalanche, in Fig. 5 we report results for larger numbers of seed excitations and a fixed excitation time of $70\,\mathrm{\mu s}$. Intuitively one expects for large seed numbers, each seed will start its own avalanche, up to the point where the seeds are so close together that no further facilitated excitations are possible. This interpretation is confirmed by Fig. 5, where $\langle N_{fac}\rangle=\langle N_{obs}\rangle-\langle N_{seed}\rangle$, i.e., the number of facilitated excitations, is plotted as a function of the number of seeds. Clearly, $\langle N_{fac}\rangle$ decreases sharply beyond around 10 seed excitations, for which the mean distance between seeds is around $2 r_{fac}$ (for the detuning $\Delta/2\pi=+24 \,\mathrm{MHz}$ used in Fig. 5, $ r_{fac}=5.7\,\mathrm{\mu m} $).

\indent Finally, as in our previous works~\cite{Malossi:14} we used the full counting statistics of the Rydberg excitations in order to obtain further insight into the underlying excitation dynamics. The experimental results, shown in \ref{1DHisto}, are discussed in detail in the next section.
\begin{figure}[htbp]
\begin{center}
\includegraphics[width=7cm]{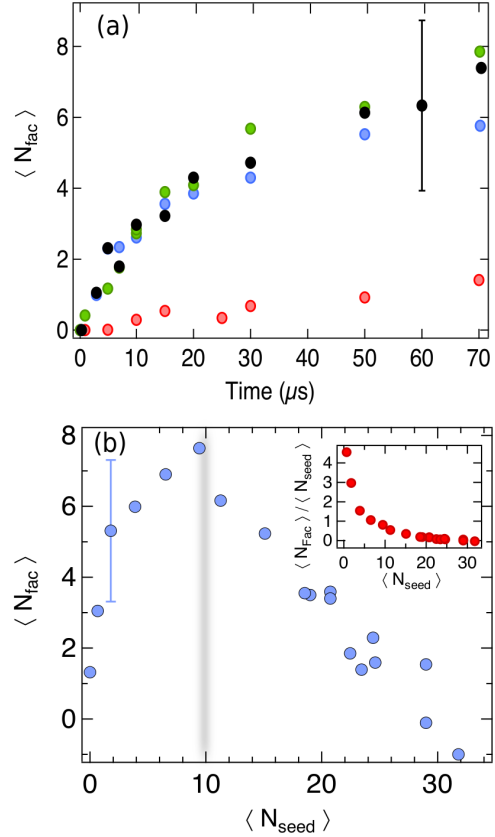}
\caption{$\langle N_{fac}\rangle$  vs the (off-resonant) excitation time and vs the number of seeds in a quasi-1D geometry. (a) $\langle N_{fac}\rangle$ vs the excitation time for values of $\langle N_{seed}\rangle$. Green data points $\langle N_{seed}\rangle=1.6$, blue points $\langle N_{seed}\rangle=3.4$, black points $\langle N_{seed}\rangle=8.6$, and red points for $\langle N_{seed}\rangle=18$. For $\langle N_{seed}\rangle=18$ the mean distance between seeds is 6.8 $\mu$m and, therefore, much smaller than $2 r_{fac}$. Parameters: $\Delta/2\pi=+24 \,\mathrm{MHz}$,  $N=490,000$ atoms (of which $1,300$ in the interaction volume) in a MOT with dimensions $160\,\mathrm{\mu m} \times 130\,\mathrm{\mu m}\times 100\,\mathrm{\mu m}$. (b) $\langle N_{fac}\rangle$ vs $\langle N_{seed}\rangle$  for a fixed excitation time of $70\,\mathrm{\mu s}$ with $\Delta/2\pi=+30 \,\mathrm{MHz}$. The typical error bar shown in the plot reflects the standard deviation. The grey line at $\langle N_{seed}\rangle\simeq10$ indicates a mean distance equal to $2 r_{fac}$. Data for $N=800,000$ atoms (of which $2,100$ in the interaction volume) in a MOT with the same dimensions as in (a). The inset reports the number of facilitated excitations per seed vs $\langle N_{seed}\rangle$.}
\label{Fig5-6}
\end{center}
\end{figure}

\begin{figure}[htbp]
\begin{center}
\includegraphics[width=9.3cm]{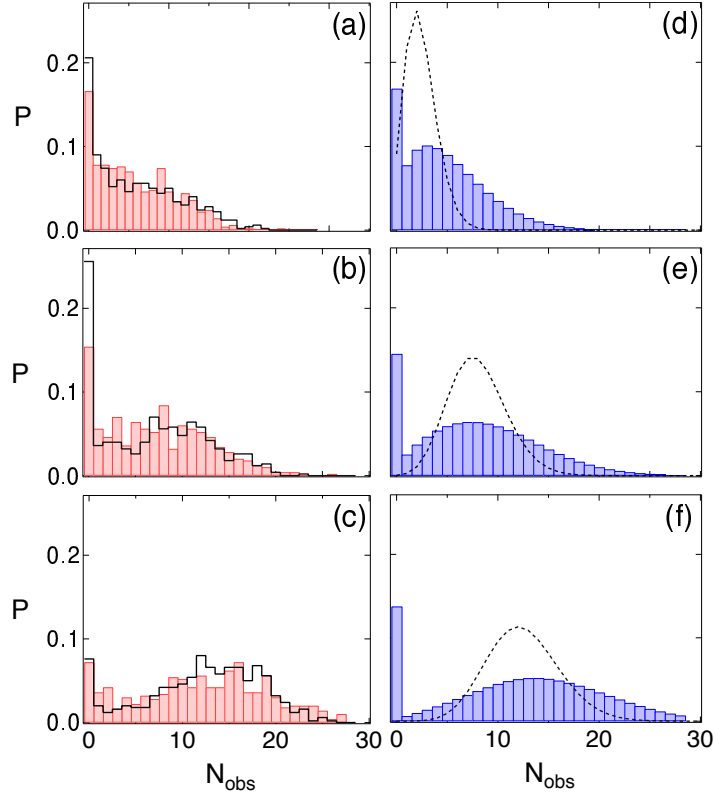}
\caption{Experimental (left) and theoretical (right) full counting distributions $ N_{obs}$ for a 1D geometry (with co-propagating laser beams). The  histograms in each row have the same mean number of atoms in the facilitation shell (at the centre of the cloud): from the top,  $\langle N_{fac}\rangle =1.3$, $1.5$  and $1.8$, respectively (between the two sets of histograms in (a)-(c), the values of $\langle N_{fac}\rangle$ matched to within $0.1$). In the experiments, at $t=0$ a mean number of $0.7$ (observed) seed excitations is created, and $\langle N_{fac}\rangle$ is obtained by adjusting the laser detuning and the atomic density using the relation $\langle N_g( r_{fac})\rangle \propto \rho/\Delta^{7/6}$. The MOT contained $2,900,000$ atoms and had dimensions $230\,\mathrm{\mu m} \times 130\,\mathrm{\mu m}\times 170\,\mathrm{\mu m}$. In (a)-(c), for the red bars $\rho=3.6\times 10^{11}\,\mathrm{cm^{-3}}$, $\rho=4.4\times 10^{11}\,\mathrm{cm^{-3}}$ and $\rho=5.3\times 10^{11}\,\mathrm{cm^{-3}}$ at fixed $\Delta/2\pi=+12$ MHz (red bars), while for the black lines $\Delta/2\pi=+15$ MHz, $\Delta/2\pi=+13.5$ MHz and $\Delta/2\pi=+12$ MHz at fixed $\rho=6.0\times 10^{11}\,\mathrm{cm^{-3}}$.(d)-(f) Theoretical counting distributions based on the simulation described in the text for different values of $\langle N_g( r_{fac})\rangle$; in each row the black dashed curve shows a Poissonian distribution with the same mean number of excitations.}
\label{1DHisto}
\end{center}
\end{figure}

\section{Analytical models and numerical simulations}

Following the analysis introduced in~\cite{Malossi:14} we model our experimental results of Fig. 4 using a simple bimodal approach described by the following   probability distribution $P(N)$:
\begin{equation}
P(N)=\alpha\delta(N- N_1) +(1-\alpha)\delta(N-N_2),
\end{equation}
where for a given number of seeds  $\langle N_{seed}\rangle/\eta$,  the quantity  $\alpha$ is the probability of having no seed
\begin{equation}
\alpha=e^{-\frac{\langle N_{seed}\rangle}{\eta}},
\end{equation}

and $N_1$ and $N_2$ represent the number of Rydberg excitations for the two modes of the model. The basic assumption is that in the absence of a seed at $t=0$ the number of excitations in the system will be $N_1$, whereas when a seed is created the successive facilitation processes lead to a final number $N_2$ of excitations.
For the above distribution $P(N)$  the  mean number  and the Mandel $Q$-parameter  are
\begin{eqnarray}
\langle N\rangle&= \frac{\langle N_{obs} \rangle}{\eta}= \alpha N_1 +(1-\alpha)N_2, \nonumber \\
Q&=\frac{Q_{obs}}{\eta}=\frac{\alpha (\langle N\rangle- N_1)^2+(1-\alpha)(\langle N \rangle- N_2)^2}{\langle N\rangle}-1.
\end{eqnarray}
\noindent We use these expressions to reproduce the dependence of $\langle N_{obs} \rangle$ and $Q_{obs}$ on $N_{seed}$ reported in Fig. \ref{Bimodal} by using reasonable values for $N_1$ and $N_2$.  The agreement between the experiment and the model is good.  The inset of Fig. 4 reports the probability distribution $P(\tilde{N})$ of the bimodal model with and without a convolution with a binomial distribution reflecting the finite detection efficiency $\eta$. This distribution reproduces the general shape of the  experimental probability distributions  of Fig. \ref{1DHisto}.\\
\indent Whilst the above model is a good approximation for the  atomic density regime where the facilitation "steps" starting from the seed occur with high probability, meaning that the mean number of atoms in each facilitation shell is much larger than unity, it is less suitable for describing our system at lower and non-uniform atomic densities. In fact, the spatial density distribution is expected to be significantly influence the avalanche process.\\
\indent Furthermore, the above model assumes a counting distribution rather than predicting it. For a 1D geometry, a simple model that predicts the counting distributions taking into account the density variation across the cloud can be derived as follows. Assuming that a seed is created at the centre of the atomic density distribution, the probability $P(s)$ that the system will realize exactly $s$ facilitation steps can be calculated by multiplying, at each step, the probabilities of having at least one atom in the $j$-th facilitation shell. This product is performed up to shell $s$, and the result is multiplied by the probability that the avalanche stops there, i.e., that there is no atom in the next shell $s+1$. This can be expressed mathematically in the following formula:
\begin{equation}
P(s)=exp\left[-\langle N_g\left((s+1) r_{fac}\right)\rangle\right] \prod_{j=1}^{s} \left(1 - exp\left[-\langle N_g(jr_{fac})\rangle\right]\right) = P(N+1)
\label{1Dprobability}
\end{equation}
\noindent where $\langle N_g(j  r_{fac})\rangle$ is the mean number of ground state atoms in the facilitation shell at distance $ j r_{fac}$ from the centre of the density distribution.
The number of steps $s=t\Gamma$ to be used in this model is determined by the total pulse time $t$ and the resonant transition rate $\Gamma$ corresponding  to the  timescale  $\tau=1/\Gamma$ for each facilitation step.\\
\indent  For the 1D geometry the full counting distribution is obtained from Eq.~(\ref{1Dprobability}) assuming that  from the central seed the avalanche starts in both directions, with the number of ground state atoms within the $j$th shell given by Eq.~(\ref{1Dfacilitation}). De-excitation events of the atoms are neglected in this model. The above process can also be simulated numerically by placing a single seed at the center of the MOT and  performing the $j$th excitation with probability $P (j + 1) = 1 - e^{\langle N_g (jr_{fac})\rangle}$. When the excitation probability reaches $\approx 1\times10^{-4}$, the simulation is stopped and the total number of excitations is recorded. The results obtained by several simulations, multiplied by the average probability $(1-\alpha)$ for the creation of at least one seed and convoluted with a binomial distribution in order to take into account our finite detection efficiency, are reported as the blue histograms of Fig \ref{1DHisto}(d-f). They demonstrate the good agreement between the mathematical model and the experimental results.\\
\indent In the above analysis we use $\langle N_g( r_{fac})\rangle$ as the only defining parameter of our system. In fact, although the number of atoms in a facilitation shell is determined by the atomic density $\rho$ as well as the radius $r_{fac}$ and the width $\delta r_{fac}$ of the facilitation shell, which depend on $\Delta$, in our model only the product of the shell volume and the density matters. This means that if we change $\Delta$ and $\rho$ keeping $\langle N_g( r_{fac})\rangle$ constant, we expect to see the same counting distributions. This feature is very well reproduced by the experimental results shown in  Fig.~\ref{1DHisto} (a) to (c), where in each case $\Delta$ and $\rho$ are chosen (red bars and black line) such as to keep $\langle N_g( r_{fac})\rangle$ (evaluated at the centre of the cloud) constant.\\
\indent The experimental and theoretical results of Fig.~\ref{1DHisto} exhibit three different types of distributions. In (a,d) the value of $\langle N_g( r_{fac})\rangle$ at the centre of the cloud is small so that even the first few facilitation steps are unlikely to occur and successive ones even more so, resulting in an exponentially decaying counting distribution;  in (b,e) $\langle N_g( r_{fac})\rangle$  is larger, leading to flat counting distributions. There the probability of creating the first facilitated excitation is less than 100 percent but still sufficient to ensure that a large number of facilitation steps is realized in most evolutions. Finally in (c,f) the bimodal distribution with one mode centered around $N_{obs}\approx 0$ and the other around $N_{obs}\approx 12$ arises for large values of $\langle N_g( r_{fac})\rangle$, confirming our interpretation of Section 4 according to which every time a seed is created, the number of excitations reaches the saturation value. In the presence of a seed, the atomic system easily realizes the first few facilitation steps with a  stop in the facilitation process when the density becomes too small to ensure the presence of an atom in the next facilitation shell. \\
\indent For both of the above models the mean number of seeds plays a role only insofar as it determines the probability of starting the avalanche process. On the other hand, in those cases where more than one seed is created one expects each seed excitation to start its own avalanche process, resulting in a faster initial growth of the number of Rydberg excitations. However, the finite volume of the MOT implies that those avalanche processes will soon start running into each other, so that above a certain number of seeds each seed excitation will only create a small number of facilitated excitations. Eventually, when the mean distance between the seed excitations is less then $\approx 2 r_{fac}$, there will not be any ground state atoms in the MOT that fulfil the facilitation condition. To verify this interpretation we have performed the avalanche excitation process in a quasi-1D geometry and in a regime of large $\langle N_{seed}\rangle$ up to $\approx 30$. Fig. 5(b) confirms this interpretation: for $\langle N_{seed}\rangle>10$, which corresponds to a mean distance between excitations below $2r_{fac}$, the number of facilitated excitations decreases towards 0.


\section{Conclusions}
In summary, we have shown that off-resonant excitation of a cloud of cold atoms to Rydberg states leads to a facilitation processes that strongly depends on the number of ground state atoms in the facilitation shell. This process is a striking example of a many body system where strong interactions lead to strongly correlated dynamics. The seed technique allows us to control the probability and the starting point of this process. We found that while a simple bimodal approach reproduces the essential features for the mean number of excitations and the Mandel Q parameter, a more refined analytical model describes the full counting distribution of the Rydberg excitation process. Our results pave the way towards flexible experimental tools for exploring Ising-like models and spin systems. For instance, by controlling the energy of the Rydberg states through light shifts or applied electric field, the spatial position of the seeds as well as the directions in which facilitation occurs could be controlled. \\

\section{Acknowledgments}
The authors acknowledge support by the European Union H2020 FET Proactive project RySQ (grant N. 640378) and by the EU Marie Curie ITN COHERENCE (Project number 265031), and thank I. Lesanovsky for discussions.

\section*{References}
\bibliography{Bibliography_SeedV10}

\end{document}